\documentclass[sigconf]{acmart}

\usepackage{booktabs} 

\setcopyright{rightsretained}

\acmDOI{10.475/123_4}

\acmISBN{123-4567-24-567/08/06}

\acmConference[WOODSTOCK'97]{ACM Woodstock conference}{July 1997}{El
  Paso, Texas USA} 
\acmYear{1997}
\copyrightyear{2016}

\acmPrice{15.00}

\begin{document}
\title{Discovering demographic data of users\\ from the evolution of their spatio-temporal entropy}


\author{Arielle Moro}
\affiliation{%
  \institution{Dep. of Information Systems, UNIL}
  \postcode{1015}
  \city{Lausanne} 
  \country{Switzerland} 
}
\email{arielle.moro@unil.ch}

\author{Beno\^{i}t Garbinato}
\affiliation{%
  \institution{Dep. of Information Systems, UNIL}
  \postcode{1015}
  \city{Lausanne} 
  \country{Switzerland} 
}
\email{benoit.garbinato@unil.ch}

\author{Val\'erie Chavez-Demoulin}
\affiliation{%
  \institution{Department of Operations - UNIL}
  \postcode{1015}
  \city{Lausanne} 
  \country{Switzerland} 
}
\email{valerie.chavez@unil.ch}

\renewcommand{\shortauthors}{A. Moro, B. Garbinato and V. Chavez-Demoulin}

\begin{abstract}

Inferring information related to users enables to highly improve the quality of many mobile services.
For example, knowing the demographic characteristics of a user allows a service to display more accurate information.
According to the literature, various works present models to detect them but, to the best of our knowledge, no one is based on the use of the spatio-temporal entropy and introduces Generalized Additive models (GAMs) in this context to reach this goal.
In this preliminary work, we present a new approach including these two key elements.
The spatio-temporal entropy enables to capture the regularity of the mobility behavior of a user, while GAMs help to predict her demographic data based on several co-variables including the spatio-temporal entropy.
The preliminary results are very encouraging to do future work since we obtain a prediction accuracy of 87\% about the prediction of the working profile of users.

\end{abstract}

%
%



\maketitle

\section{Introduction}\label{sec:introduction}

In recent years, discovering demographic characteristics of users has caught attention of the research community and companies.
Existing services can indeed highly improve the quality of the services by displaying accurate information to the users.
However, these services do not always know the characteristics of users and need to infer them by exploring other related information.
A first work demonstrates that it is possible to guess attributes of users based on their social network, by analyzing friends and communities more specifically~\cite{Mislove:2010:YYK:1718487.1718519}.
In a second work~\cite{ying2012demographic}, Ying et al. propose a model, called \emph{Multi-Level Classification Model}, that combines several classification models in order to infer demographic user attributes.
The global model takes~45 features into account that describe a user behavior.
Then, it has been shown that communication records are also very interesting and powerful to detect demographic data about users~\cite{Dong:2014:IUD:2623330.2623703}.
Fourthly, Zhong et al. \cite{Zhong:2015:YYG:2684822.2685287} prove that exploring location check-ins gives a lot of information about user profiles.
Finally, the authors of~\cite{Wang:2017:IDS:3041021.3054140} create a technique helping to find demographic data of users by analyzing AP-trajectories.
These last two papers highlight the importance of user's movements.
However, they do not explore the characteristics of her movements from a spatio-temporal entropy perspective.
This latter can reveal crucial information, such as the regularity as well as the spatio-temporal uncertainty of the user's movements.
Song et al. already proved that the entropy is a good measure to assess the level of predictability of a user as indicated in~\cite{song2010limits}.
We assume that the spatio-temporal entropy can also help to discover demographic data of a user because the entropy evolution of users having the same demographic characteristic may be similar.
In this work and to the best of our knowledge, we present a novel approach to discover demographic data of users based on their spatio-temporal entropy and Generalized Additive Models (GAMs).
The strength of a GAM is that this model is not only helpful to provide predicted values but also to give a good understanding of the influence of the co-variables used on the response variable.
GAM also provides a framework allowing formal inference, that is, formal significance testing of the covariates. 
Finally, as a flexible extension of the standard Generalized Linear Models (GLMs), the influence of the covariates is not constrained to have parametric forms. 
If we take the following simple example, a GAM can highlight that a high entropy is more linked with an age-group of young users.
Consequently, if the GAM input is an high entropy value, the predicted age value may represent a young user with a certain uncertainty provided by confidence interval.
Moreover, we show preliminary results obtained with real mobility traces from two datasets containing real mobility traces including raw locations captured via GPS, WPS and radio cells.
Finally, we conclude with the main preliminary finding and the most promising lines of our future work.
\section{Approach}\label{sec:methodology}

We present our approach to solve the discovery of demographic data of users in three steps: computing the spatio-temporal entropy sequence of a user, building GAMs and evaluating the predicted values from the fitted GAMs.

\subsection{Computing spatio-temporal entropy sequence}
\begin{figure}[t!]
\centering
\includegraphics[scale=0.35]{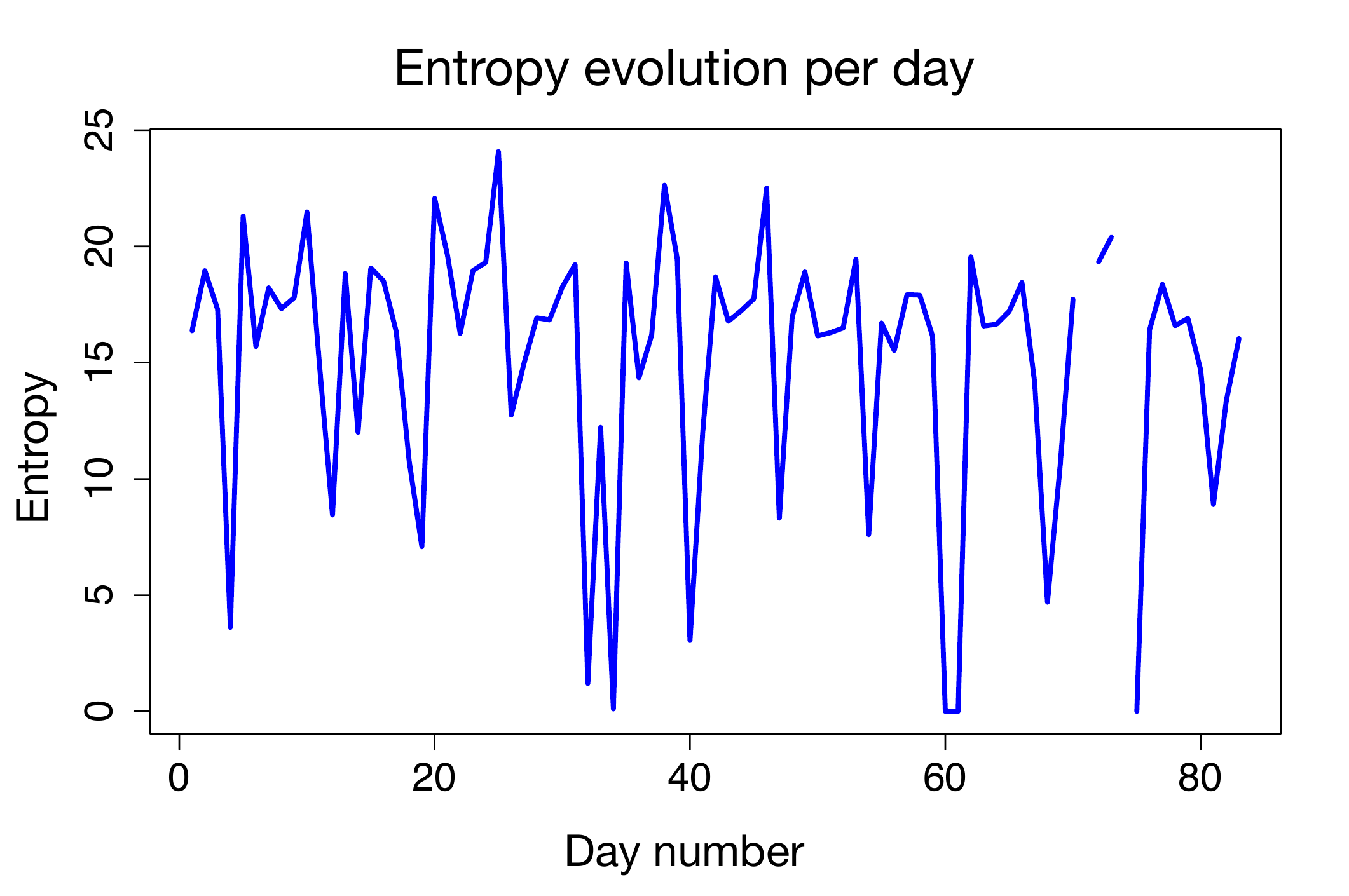}
\caption{Evolution of the spatio-temporal entropy per day}
\label{fig:entropyEvolution}
\end{figure}

In order to compute the spatio-temporal entropy sequence of a user based on her location history, we firstly need to create a spatial grid to discretize the space in which the user is moving.
The space is divided into a 2-dimensional plane in which the position of each cell in this grid is described as a pair $(i,j)$.
We also note $n$ and $m$ the numbers of cells along the x-axis and along the y-axis respectively.
We divide the time duration of the location history of the user into $T$ time slices having the same period of time, e.g., 86400 seconds (one day).
Then, we compute the time proportion $p_{i,j}^t$ for each visited cell of the grid by the user, where $t$ is a specific time slice.
Finally, the entropy computation of a specific time slice is detailed in Equation~\ref{equ:entropyOfATimeSlice}.
This latter indicates that the entropy is a percentage ranged between 0\% and 100\%.
At the end of this process, we must obtain a sequence of $T$ entropies containing the movement rhythm of the user as described in Figure~\ref{fig:entropyEvolution} (a) obtained with a very rich private dataset of an individual.\\

\begin{equation}
	entropy_{t} = - \frac{\sum\limits_{i=1}^n \sum\limits_{j=1}^m p_{i,j}^t \log_{} p_{i,j}^t}{\log_{} n \times m} \times 100.0
\label{equ:entropyOfATimeSlice}
\end{equation}


\subsection{Creating Generalized Additive Models (GAMs)}
In order to infer a demographic variable (e.g., gender, age group or working profile), we must fit a GAM for each demographic variable.
Each GAM is trained with multiple user datasets from which we know the value of the demographic variable we want to infer but also their spatio-temporal entropy sequence as well as other co-variables interesting to add to improve the quality of the predictions.
All the co-variables must have a value for each time slice analyzed, which can be the same in the case of the age group variable or different if it is the entropy.
We can select different types of co-variables, such as spatial co-variables (e.g., maximum distance travelled during one day) as well as temporal co-variables (e.g., day number of the week).\\


\subsection{Evaluating the performance of the GAM}
In order to assess the accuracy of the prediction of the demographic variable of a sample of test  users, we train the GAM corresponding to a demographic variable with 90\% of the users of a dataset.
Then, we use the values of each new user of the 10\% of the remaining users to do the prediction process and compare the predicted value with the real demographic variable of the user.
\vspace{3.5cm}
\section{Preliminary results}\label{sec:experiments_results}

In order to obtain preliminary results, we use a dataset called~\emph{Mobile Data Challenge (MDC) Dataset} presented in~\cite{laurila2012mobile}.
Regarding the demographic data discovery, we select all users who indicate information about the three following demographic variables: gender, age group and working profile, i.e., 153 users in total.
Table~\ref{tab:demographic_group_discovery_results} describe the prediction accuracy results obtained for the three demographic variables.
Interestingly and somehow not surprisingly, it is easier to predict the age group variable and the working profile than the gender of a user.


\begin{table}[t!]
  \footnotesize
  \caption{Demographic variable prediction accuracy results}
  \label{tab:demographic_group_discovery_results}
  \begin{tabular}{ccl}
    \toprule
    Demo. variable & Pred. accuracy & Description of groups\\
    \midrule
    Gender & 60\% & Female/Male \\
    Age group & 73\% & <22 / >=22 \& <33 / >=33 years old\\
    Working profile & 87\% & Full time/Part time/Other working profiles\\
    \bottomrule
  \end{tabular}
\end{table}
\section{Conclusion and future work}\label{sec:conclusion_futurework}

We presented an approach to discover demographic characteristics of users.
The main finding is that entropy seems a good measure to be used through flexible GAM fitting to discriminate demographic characteristics.
Regarding the future work, we will firstly analyze the impact of the addition of other types of co-variables on the demographic data discovery, such as climatic variables (e.g., temperature of the day), communication variables (e.g., number of calls of the day) and probably much more.
Then, we will compare the demographic accuracy results obtained with GAMs with other models, such as classifiers.
Finally, we will use very rich user datasets, in terms of location frequency and precision, in order to deeper study user entropy rhythms and do more evaluations about demographic profiles at a fine grained level, e.g., at a hourly basis.

\bibliographystyle{plain}
\bibliography{sigproc}

\begin{thebibliography}{1}

\bibitem{Dong:2014:IUD:2623330.2623703}
Yuxiao Dong, Yang Yang, Jie Tang, Yang Yang, and Nitesh~V. Chawla.
\newblock Inferring user demographics and social strategies in mobile social
  networks.
\newblock In {\em Proceedings of the 20th ACM SIGKDD International Conference
  on Knowledge Discovery and Data Mining}, KDD '14, pages 15--24, New York, NY,
  USA, 2014. ACM.

\bibitem{laurila2012mobile}
Juha~K Laurila, Daniel Gatica-Perez, Imad Aad, Olivier Bornet, Trinh-Minh-Tri
  Do, Olivier Dousse, Julien Eberle, Markus Miettinen, et~al.
\newblock The mobile data challenge: Big data for mobile computing research.
\newblock In {\em Pervasive Computing}, number EPFL-CONF-192489, 2012.

\bibitem{Mislove:2010:YYK:1718487.1718519}
Alan Mislove, Bimal Viswanath, Krishna~P. Gummadi, and Peter Druschel.
\newblock You are who you know: Inferring user profiles in online social
  networks.
\newblock In {\em Proceedings of the Third ACM International Conference on Web
  Search and Data Mining}, WSDM '10, pages 251--260, New York, NY, USA, 2010.
  ACM.

\bibitem{song2010limits}
Chaoming Song, Zehui Qu, Nicholas Blumm, and Albert-L{\'a}szl{\'o}
  Barab{\'a}si.
\newblock Limits of predictability in human mobility.
\newblock {\em Science}, 327(5968):1018--1021, 2010.

\bibitem{Wang:2017:IDS:3041021.3054140}
Pinghui Wang, Feiyang Sun, Di~Wang, Jing Tao, Xiaohong Guan, and Albert Bifet.
\newblock Inferring demographics and social networks of mobile device users on
  campus from ap-trajectories.
\newblock In {\em Proceedings of the 26th International Conference on World
  Wide Web Companion}, WWW '17 Companion, pages 139--147, Republic and Canton
  of Geneva, Switzerland, 2017. International World Wide Web Conferences
  Steering Committee.

\bibitem{ying2012demographic}
Josh Jia-Ching Ying, Yao-Jen Chang, Chi-Min Huang, and Vincent~S Tseng.
\newblock Demographic prediction based on users mobile behaviors.
\newblock {\em Mobile Data Challenge}, 2012.

\bibitem{Zhong:2015:YYG:2684822.2685287}
Yuan Zhong, Nicholas~Jing Yuan, Wen Zhong, Fuzheng Zhang, and Xing Xie.
\newblock You are where you go: Inferring demographic attributes from location
  check-ins.
\newblock In {\em Proceedings of the Eighth ACM International Conference on Web
  Search and Data Mining}, WSDM '15, pages 295--304, New York, NY, USA, 2015.
  ACM.

\end{thebibliography}

\end{document}